\documentclass[runningheads,a4paper]{llncs}

\usepackage{amsfonts}
\usepackage{dsfont}

\usepackage{mathtools}

\usepackage{amssymb}

\usepackage{amsmath}

\usepackage{amsthm}

\usepackage{graphicx}

\usepackage{epstopdf}

\usepackage[hidelinks]{hyperref}
\usepackage{subfig}
\usepackage{caption}

\usepackage{float}

\makeatletter

\def\holdocspecials{\do\ \do\$\do\&%
  \do\#\do\^\do\^^K\do\_\do\^^A\do\%}

\def\holtt{\trivlist \item[]\if@minipage\else\vskip\parskip\fi
\leftskip\@totalleftmargin\rightskip\z@
\parindent\z@\parfillskip\@flushglue\parskip\z@
\@tempswafalse \def\par{\if@tempswa\hbox{}\fi\@tempswatrue\@@par}
\obeylines \tt \let\do\@makeother \holdocspecials
 \frenchspacing\@vobeyspaces}

\makeatother

\newlength{\hsbw}
\newcommand\HOLSpacing{13pt}

   \newcommand\hilbert{\varepsilon}

   \newcommand{\Cond}{\(\Longrightarrow\)}
   \newcommand{\Eqv}{\(\equiv\)}
   \newcommand{\Iff}{\(\Leftrightarrow\)}
   \newcommand{\Fa}{\(\scriptsize \forall\)}
   \newcommand{\Et}{\(\exists\)}
   \newcommand{\Eu}{\(\exists_{unique}\)}

   \newcommand{\Func}{\(\to\)}
   \renewcommand{\Bar}{\(\mid\)}

   \newcommand{\Lam}{\(\lambda\)}
   \newcommand{\Plus}{\(+\)}
   \newcommand{\Minus}{\(-\)}
   \newcommand{\Prime}{\('\)}
   \newcommand{\Und}{\_}
   \newcommand{\Lt}{\(<\)}
   \newcommand{\Gt}{\(>\)}
   \newcommand{\Leq}{\(\leq\)}
   \newcommand{\Geq}{\(\geq\)}
   \newcommand{\Eq}{\(=\)}

\newcommand{\Hilbert}{\(\hilbert\)}
\newcommand{\Imp}{\(\Rightarrow\)}
\newcommand{\Conj}{\(\wedge\)}
\newcommand{\Disj}{\(\vee\)}
\newcommand{\Neg}{\(\neg\)}
\newcommand{\Pnd}{\(\Diamond\)}

\long\def\rechol#1#2#3{\let\next=\rechol\def\postnext{#2#3}\ifx#1\end
\let\next=\relax\def\postnext{\relax}
\else\ifx#1!\Fa                                          
\else\ifx#1@\Hilbert                                     
\else\ifx#1\#\Pnd                                        
\else\ifx#1'\Prime                                       
\else\ifx#1~\Neg                                         
\else\ifx#1\~\Neg
\else\ifx#1_\Und                                         
\else\ifx#1+\Plus
\else\ifx#1\/\Disj                                       
\else\ifx#1\.\Lam                                        
\else\ifx#1>\ifx#2=\Geq\def\postnext{#3}\else\Gt\fi      
\else\ifx#1?\ifx#2!\Eu\def\postnext{#3}\else\Et\fi       
\else\ifx#1-\ifx#2\>\Func\def\postnext{#3}               
	    \else\Minus\fi				 
\else\ifx#1|\ifx#2-\Turns\def\postnext{#3}\else\Bar\fi   
\else\ifx#1<\ifx#2=\ifx#3>\Iff\def\postnext{}            
                   \else\Leq\def\postnext{#3}\fi         
            \else\Lt\fi
\else\ifx#1=\ifx#2=\ifx#3>\Imp\def\postnext{}            
                   \else\Eqv\def\postnext{#3}\fi         
            \else\ifx#2>\Cond\def\postnext{#3}
                 \else\Eq\fi\fi
\else\ifx#1/\ifx#2\^^M\Conj\par\def\postnext{#3}         
            \else\ifx#2\ \Conj\ \def\postnext{#3}\else#1\fi\fi  
\else#1\fi\fi\fi\fi\fi\fi\fi\fi\fi\fi\fi\fi\fi\fi\fi\fi\fi\fi
\expandafter\next\postnext}

\usepackage{epsfig}

\usepackage{array,longtable}

\usepackage{mdframed}

\usepackage{multirow}

\usepackage{url}

\PassOptionsToPackage{hyphens}{url}\usepackage{hyperref}

\urldef{\mailsa}\path|{adnan.rashid, osman.hasan}@seecs.nust.edu.pk|

\newcommand{\keywords}[1]{\par\addvspace\baselineskip
\noindent\keywordname\enspace\ignorespaces#1}

\begin{document}

\mainmatter  

\title{Formal Analysis of Linear Control Systems using Theorem Proving}

\titlerunning{Formal Analysis of Linear Control Systems}

%
%

\author{Adnan Rashid \and Osman Hasan}

%
\authorrunning{A. Rashid and O. Hasan}


\institute{School of Electrical Engineering and Computer Science (SEECS)\\
National University of Sciences and Technology (NUST)\\
Islamabad, Pakistan\\
\mailsa\\
}

%
%
\maketitle

\begin{abstract}

Control systems are an integral part of almost every engineering and physical system and thus their accurate analysis is of utmost importance. Traditionally, control systems are analyzed using paper-and-pencil proof and computer simulation methods, however, both of these methods cannot provide accurate analysis due to their inherent limitations.
Model checking has been widely used to analyze control systems but the continuous nature of their environment and physical components cannot be truly captured by a state-transition system in this technique.
To overcome these limitations, we propose to use higher-order-logic theorem proving for analyzing linear control systems based on a formalized theory of the Laplace transform method. For this purpose, we have formalized the foundations of linear control system analysis in higher-order logic so that a linear control system can be readily modeled and analyzed. The paper presents a new formalization of the Laplace transform and the formal verification of its properties that are frequently used in the transfer function based analysis to judge the frequency response, gain margin and phase margin, and stability of a linear control system. We also formalize the active realizations of various controllers, like Proportional-Integral-Derivative (PID), Proportional-Integral (PI), Proportional-Derivative (PD), and various active and passive compensators, like lead, lag and lag-lead. For illustration, we present a formal analysis of an unmanned free-swimming submersible vehicle using the HOL Light theorem prover.

\end{abstract}

\section{Introduction} \label{SEC:Intro}

Linear control systems are widely used to regulate the behavior of many safety-critical applications, such as process control, aerospace, robotics and transportation. The first step in the analysis of a linear control system is the construction of its equivalent mathematical model by using the physical and engineering laws. For example, in the case of electrical systems, we need to model the currents and voltages passing through the electrical components and their interactions in the corresponding electrical circuit using the system governing laws, such as Kirchhoff's current law (KCL) and Kirchhoff's voltage law (KVL). The mathematical model is then used to derive differential equations describing the relationship between the inputs and outputs of the underlying system. The next step in the analysis of a linear control system is to solve these equations to obtain a transfer function, which is in turn used to assess many interesting control system characteristics, such as frequency response, phase margin and gain margin. However, solving these equations in the time domain is not so straightforward as they usually involve the integral and differential operators. The Laplace transform, which is an integral based transform method, is thus often used to convert these differential equations to their equivalent algebraic equations in $s$-domain by converting the differential and integral operations into multiplication and division operators, respectively. This algebraic equation can be quite easily solved to obtain the corresponding transfer function, frequency response, gain margin and the phase margin and perform the stability analysis of the given control system.

Traditionally, the linear control system analysis is performed using paper-and-pencil proof methods. However, these methods are human-error prone and cannot be relied upon for the analysis of safety-critical applications. Moreover, there is always a risk of misusing an existing mathematical result as this manual analysis method does not provide the assurance that a mathematical law would be used only if all of its required assumptions are valid. Computer simulation and numerical methods are also frequently used to analyze linear control systems. However, they also compromise the accuracy of the results due to the involvement of computer arithmetic and the associated round-off errors. Computer algebra systems (CAS), such as Mathematica~\cite{lutovac2006symbolic}, are also used for the Laplace transform based analysis of linear control systems. However, CAS are primarily based on unverified symbolic algorithms and thus there is no formal proof to ascertain the accuracy of their analysis results. Given the inaccurate nature of all the above-mentioned analysis techniques, they are not very suitable to analyze control systems used in safety-critical domains, where even a slight error in analysis may lead to disastrous consequences, including the loss of human lives.

To overcome the above-mentioned limitations, model checking~\cite{hasan2015formal} has been also used to analyze control systems~\cite{johnson2002model,tiwari2002series} but the continuous nature of their environment and physical components cannot be truly captured by a state-transition system in this technique.
Similarly, a Hoare logic based framework~\cite{boulton2003hoare} and the KeYmaera tool~\cite{arechiga2012using} have been used for the formal frequency domain analysis and verification of the safety properties of control systems with sampled-time controllers, respectively. However, the former is limited to the analysis of systems that can be expressed using a block diagram with a tree structure, whereas in the later, the continuous nature of the models is abstracted in the formal modeling process and hence the completeness of the analysis is compromised in both cases.

Recently, the HOL Light theorem prover has been used for the formal analysis of control systems.
\textit{Hasan et al.} presented a formalization of the block diagrams in HOL Light and used it to reason about the transfer function and the steady-state error analysis of a feedback control system~\cite{hasan2013formal}. \textit{Ahmed et al.} used this formalization of block diagrams to verify the steady-state error of a unity feedback control system~\cite{ahmad2014formal}. Similarly, \textit{Beillahi et al.} formalized the signal flow graphs in HOL Light, which can be used to formally verify transfer functions of linear control systems~\cite{beillahi2015formal}.
However, all these existing works focus on the verification of the transfer functions for a control system and, to the best of our knowledge, no prior work dealing with the formal analysis of dynamics of a linear control system exists in the literature of higher-order-logic theorem proving.

\begin{figure}[h]
\centering
\scalebox{0.20}
{\includegraphics[trim={5.0 0.4cm 5.0 0.4cm},clip]{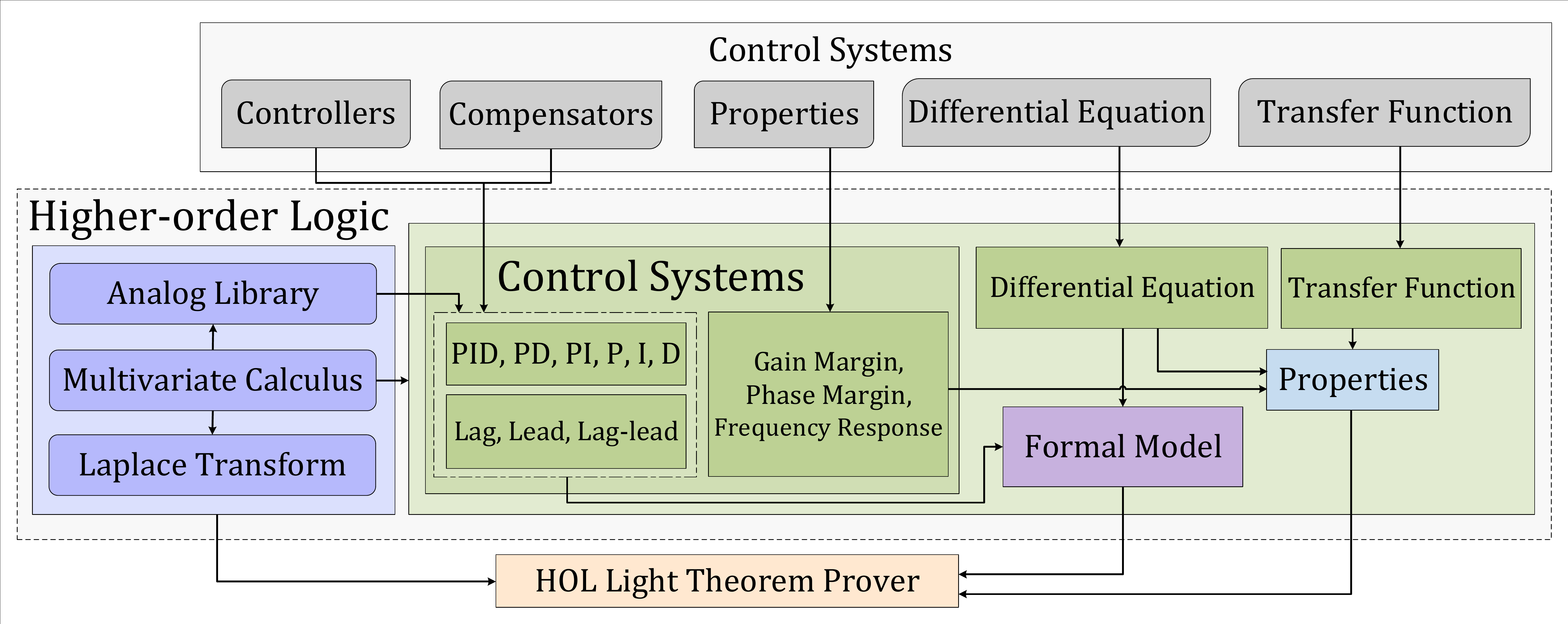}}
\caption{Proposed Framework}
\label{FIG:proposed_framework}
\end{figure}

In this paper, we present a framework to conduct the formal analysis of dynamical characteristics of a linear control system using higher-order-logic theorem proving.
The main idea behind the proposed framework, depicted in Fig.~\ref{FIG:proposed_framework}, is to formalize all the foundational components of a linear control system to facilitate formal modeling and reasoning about linear control systems within the sound core of a theorem prover.
For this purpose, we built upon the higher-order-logic formalizations of Multivariable calculus~\cite{harrison2013hol} and a library of analog components, like resistor, capacitor and inductor~\cite{taqdees2017tflac}.
We present a \textit{new formalization of Laplace transform}, which includes the formal verification of some of its frequently used properties in reasoning about the transfer function of an $n$-order system.
We also formalized some widely used \textit{characteristics of linear control systems}, such as frequency response, gain margin and phase margin, which can be used for the stability analysis of a linear control system. Moreover, we formalize the \textit{active realizations of various controllers}, such as Proportional-Integral-Derivative (PID), Proportional-Integral (PI), Proportional-Derivative (PD), Proportional (P), Integral (I) and Derivative (D) and various \textit{active and passive compensators}, such as lag, lead and lag-lead.

The proposed framework, depicted in Fig.~\ref{FIG:proposed_framework}, allows us to build a formal model of the given linear control system, based on the active realizations of its

\noindent  controllers and compensators, the passive realizations of compensators and differential equations. Moreover, it also allows to formalize the behavior of the given linear control system in terms of its differential equation, transfer function specification and its properties, such as phase margin, frequency response and gain margin.
We can then use these formalized models and properties to verify an implication relationship between them, i.e., model implies its specification.
In order to demonstrate the effectiveness of our proposed formalization, we formalize the control system of an unmanned free-swimming submersible vehicle~\cite{nise2007control}.
We have used the HOL Light theorem prover \cite{harrison-hol-light} for the proposed formalization in order to build upon its multivariable calculus theories.
We have also developed a tactic that can be used to automatically verify the transfer function of any control system up to $20^{th} order$.
This tactic was found to be very handy in the formal analysis of the unmanned submersible vehicle.

\section{Multivariable Calculus Theories in HOL Light} \label{SEC:Mult_cal_theories}

An N-dimensional vector is formalized in the multivariable theory of HOL Light as a $\mathds{R}^N$ column matrix of real numbers~\cite{harrison2013hol}. All of the multivariable calculus theorems are verified for functions with an arbitrary data-type $\mathds{R}^N \rightarrow \mathds{R}^M$.

A complex number is defined as a $2$-dimensional vector, i.e., a $\mathds{R}^2$ matrix.

\begin{flushleft}
\begin{definition}
\label{DEF:cx_and_ii}
{\footnotesize
 \textup{\texttt{$\vdash$ $\forall$ a. Cx a = complex (a, \&0)  \\
$\vdash$ ii = complex (\&0, \&1) }}}
\end{definition}
\end{flushleft}

\noindent \texttt{Cx} : $\mathds{R} \rightarrow \mathds{R}^2$ is a type casting function that accepts a real number and returns its corresponding complex number with the imaginary part equal to zero, where the $\texttt{\&}$ operator
type casts a natural number to its corresponding real number. Similarly, $\mathtt{ii}$ (iota) represents a complex number having
the real part equal to zero and the magnitude of the imaginary part equal to 1.

\begin{flushleft}
\begin{definition}
\label{DEF:re_im_lift_drop}
{\footnotesize
\textup{\texttt{$\vdash$ $\forall$ z. Re z = z\$1  \\  $\vdash$ $\forall$ z. Im z = z\$2 \\
$\mathtt{\ }$\hspace{-0.33cm} $\vdash$ $\forall$ x. lift x = (lambda i. x)  \\
$\vdash$ $\forall$ x. drop x = x\$1
}}}
\end{definition}
\end{flushleft}

\noindent The function $\mathtt{Re}$ accepts a complex number (2-dimensional vector) and returns its real part. Here, the notation $\mathtt{z\$i}$ represents the
$i^{th}$ component of vector $\texttt{z}$. Similarly, $\mathtt{Im}$ takes a complex number and returns its imaginary part. The
function $\mathtt{lift}$ accepts a variable of type $\mathds{R}$ and maps it to a 1-dimensional vector with the input variable as
its single component. Similarly, $\mathtt{drop}$ takes a 1-dimensional vector and returns its single element as a real number.

\begin{flushleft}
\begin{definition}
\label{DEF:exp_ccos_csine}
{\footnotesize
 \textup{\texttt{$\vdash$ $\forall$ x. exp x = Re (cexp (Cx x))
}}}
\end{definition}
\end{flushleft}

\noindent The complex exponential and real exponentials are represented as $\texttt{cexp}:\mathds{R}^2 \rightarrow \mathds{R}^2$ and
$\mathtt{exp} : \mathds{R} \rightarrow \mathds{R}$ in HOL Light, respectively.

\begin{flushleft}
\begin{definition}
\label{DEF:vector_integral}
{\footnotesize
 \textup{\texttt{$\vdash$ $\forall$ f i. integral i f = (@y. (f has\_integral y) i)
}}} \\
{\footnotesize
 \textup{\texttt{$\vdash$ $\forall$ f i. real\_integral i f = (@y. (f has\_real\_integral y) i) }}}
\end{definition}
\end{flushleft}

\noindent The function $\mathtt{integral}$ represents the vector integral and is defined using the Hilbert choice operator $\texttt{@}$ in the functional form. It takes the integrand function $\texttt{f}$, having an arbitrary type $\mathds{R}^N \rightarrow \mathds{R}^M$, and a vector-space $\mathtt{i}: \mathds{R}^N \rightarrow \mathds{B}$, which defines the region of convergence as $\mathds{B}$ represents the boolean data type, and returns a vector $\mathds{R}^M$, which is the integral of $\mathtt{f}$ on $\mathtt{i}$. The function $\mathtt{has\_integral}$ represents the same relationship in the relational form.
Similarly, the function $\mathtt{real\_integral}$ accepts the integrand function $\mathtt{f} : \mathds{R} \rightarrow \mathds{R}$ and a set of real numbers $\mathtt{i}: \mathds{R} \rightarrow \mathds{B}$ and returns the real-valued integral of function $\mathtt{f}$ over $\mathtt{i}$.
The region of integration, for both of the above integrals can be defined to be bounded by a vector interval $[a, b]$ or real interval $[a, b]$ using the HOL Light functions $\mathtt{interval \ [a,b]}$ and $\mathtt{real\_interval \ [a,b]}$, respectively.

\begin{flushleft}
\begin{definition}
\label{DEF:vector_derivative} \emph{ } \\
{\footnotesize
\textup{\texttt{$\vdash$ $\forall$f net. vector\_derivative f net = (@f'.(f has\_vector\_derivative f') net) }}}
\end{definition}
\end{flushleft}

\noindent The function $\mathtt{vector\_derivative}$ takes a function $\texttt{f} : \mathds{R}^1 \rightarrow \mathds{R}^M$ and a
$\texttt{net} : \mathds{R}^1 \rightarrow \mathds{B}$, which defines the point at which $\texttt{f}$ has to be differentiated, and
returns a vector of data-type $\mathds{R}^M$, which represents the differential of $\texttt{f}$ at $\texttt{net}$. The function
$\mathtt{has\_vector\_derivative}$ defines this relationship in the relational form.

\begin{flushleft}
\begin{definition}
\label{DEF:limit_of_function}
{\footnotesize
\textup{\texttt{$\vdash$ $\forall$ f net. lim net f = (@l. (f
$\rightarrow$ l) net) }}}
\end{definition}
\end{flushleft}

\noindent The function $\mathtt{lim}$ accepts a $\texttt{net}$ with elements of arbitrary data-type $\mathds{A}$ and a function $\texttt{f} : \mathds{A} \rightarrow \mathds{R}^M$ and returns $\texttt{l}$ of data-type $\mathds{R}^M$, i.e., the value to which $\texttt{f}$ converges at the given $\texttt{net}$.

\section{Formalization of Laplace Transform} \label{SEC:Formalization_of_Laplace}

Mathematically, Laplace transform is defined for a function $f:\mathds{R}^1 \rightarrow \mathds{C}$ as~\cite{beerends2003fourier}:

\small

\begin{equation}\label{EQ:laplace_transform}
\mathcal{L} [f(t)] = F(s) = \int_{0}^{\infty} {f(t)e^{-s t}} dt, \ s \ \epsilon \  \mathds{C}
\end{equation}

\normalsize

We formalize Equation \ref{EQ:laplace_transform} in HOL Light as follows:

\begin{flushleft}
\begin{definition}
\label{DEF:laplace_transform}
{\footnotesize
\textup{\texttt{$\vdash$ $\forall$ s f. laplace\_transform f s = \\
$\mathtt{\ }$\hspace{0.7cm} integral \{t| \&0 <= drop t\} ($\lambda$t. cexp (--(s $\ast$ Cx (drop t))) $\ast$ f t) }}}
\end{definition}
\end{flushleft}

\noindent The function \texttt{laplace\_transform} accepts a complex-valued function $ \texttt{f}: \mathds{R}^1 \rightarrow \mathds{R}^2 $
and a complex number $\texttt{s}$ and returns the Laplace transform of $ \texttt{f} $ as represented by
Equation~\ref{EQ:laplace_transform}. In the above definition, we used the complex exponential function $ \texttt{cexp}: \mathds{R}^2
\rightarrow \mathds{R}^2 $ because the return data-type of the function $ \texttt{f} $ is $ \mathds{R}^2 $. Here, the data-type of $ \texttt{t} $ is $ \mathds{R}^1 $ and to multiply it with the complex number $ \mathtt{s} $, it is first converted into a real number by
using $\texttt{drop}$ and then it is converted to data-type $ \mathds{R}^2 $ using $ \texttt{Cx} $. Next, we use the vector
function $ \texttt{integral} $ (Definition~\ref{DEF:vector_integral}) to integrate the expression $ f(t)e^{-i \omega t} $ over
the positive real line since the data-type of this expression is $ \mathds{R}^2 $. The region of integration is \texttt{\{t | \&0 <= drop t\}}, which represents the positive real line. Laplace transform was earlier formalized using a limiting process as~\cite{taqdees2013formalization}:

\begin{flushleft}
{\footnotesize
\textup{\texttt{$\vdash$ $\forall$ s f. laplace f s = lim at\_posinfinity ($\lambda$b. integral    \\
$\mathtt{\ }$\hspace{0.2cm} (interval [lift (\&0), lift b]) ($\lambda$t. cexp (--(s $\ast$ Cx (drop t))) $\ast$ f t))
}}}
\end{flushleft}


\noindent  However, the HOL Light definition of the integral function implicitly encompasses infinite limits of integration. So, our definition covers the region of integration, i.e., $[0, \infty)$, as \texttt{\small{\{t | \&0 <= drop t\}}} and is equivalent to the definition given in~\cite{taqdees2013formalization}. However, our definition considerably simplifies the reasoning process in the verification of Laplace transform properties since it does not involve the notion of limit.

The Laplace transform of a function $f$ exists, if $f$ is piecewise smooth and is of exponential order on the positive real
line~\cite{beerends2003fourier,rashid2017tmformalization}. A function is said to be piecewise smooth on an interval if it is piecewise
differentiable on that interval.

\begin{flushleft}
\begin{definition}
\label{DEF:laplace_existence}
{\footnotesize
\textup{\texttt{$\vdash$ $\forall$ s f. laplace exists f s $\Leftrightarrow$ \\
$\mathtt{\ }$\hspace{0.0cm} ($\forall$ b. f piecewise\_differentiable\_on interval [lift (\&0),lift b] ) $\wedge$ \\
$\mathtt{\ }$\hspace{0.0cm} ($\exists$ M a. Re s > drop a $\wedge$ exp\_order\_cond f M a) }}}
\end{definition}
\end{flushleft}

\noindent The function $\texttt{exp\_order\_cond}$ in the above definition represents the exponential order condition necessary for the existence of the Laplace transform~\cite{taqdees2013formalization,beerends2003fourier}:

\begin{flushleft}
\begin{definition}
\label{DEF:exp_order_condition}
{\footnotesize
\textup{\texttt{$\vdash$ $\forall$ f M a. exp\_order f M a $\Leftrightarrow$ \&0 < M $\wedge$ \\
$\mathtt{\ }$\hspace{0.6cm} ($\forall$ t. \&0 <= t $\Rightarrow$ norm (f (lift t)) <= M $\ast$ exp (drop a $\ast$ t)) }}}
\end{definition}
\end{flushleft}

We used Definitions~\ref{DEF:laplace_transform},~\ref{DEF:laplace_existence} and~\ref{DEF:exp_order_condition} to formally
verify some of the classical properties of Laplace transform, given in Table~\ref{TAB:properties of_lap_trans_new}.
The properties namely linearity, frequency shifting, differentiation and integration were already verified using the formal definition of the Laplace transform~\cite{taqdees2013formalization}. We formally verified these using our new definition of the Laplace transform. Moreover, we formally verified some new properties, such as, time shifting, time scaling, cosine and sine-based modulations and the Laplace transform of a $n$-order differential equation.
The assumptions of these theorems describe the existence of the corresponding Laplace transforms. For example, the predicate \texttt{laplace\_exists\_higher\_deriv} in the theorem corresponding to the $n$-order differential equation ensures that the La-


\begin{footnotesize}
    \begin{longtable}{|p{2cm}|p{3cm}|p{7cm}|p{3cm}|}
\caption{Properties of Laplace Transform}
\label{TAB:properties of_lap_trans_new}
\endfirsthead
\endhead
    \hline
    \hline
    \multicolumn{1}{l|}{Property}   &
    \multicolumn{1}{l}{\hspace{0.0cm} Formalized Form}

     \\ \hline \hline




    \multicolumn{1}{l|}{ {$\begin{array} {lcl} \textbf{Integrability} \\
    \hspace{0.0cm} \textit{$ e^{- s t} f(t)\ integrable\  $ } \\
\textit{$\mathtt{\ }$\hspace{0.4cm} $ on\ [0, \infty) $     }
 \end{array}$}  }  &

   \multicolumn{1}{l}{{$\begin{array} {lcl} \textup{\texttt{\hspace{0.0cm}$\vdash$ $\forall$ f s. laplace\_exists f s $\Rightarrow$    }} \\
\textup{\texttt{$\mathtt{\ }$\hspace{0.0cm} ($\lambda$t. cexp (--(s $\ast$ Cx (drop t))) $\ast$ f t)  }} \\
\textup{\texttt{$\mathtt{\ }$\hspace{2.00cm} integrable\_on \{t | \&0 <= drop t\}  }}
 \end{array}$}}    \\ \hline




   \multicolumn{1}{l|}{ {$\begin{array} {lcl} \textbf{Linearity} \\
   \hspace{0.0cm} \textit{$ \mathcal{L} [ \alpha f(t) + \beta g(t)] = $ } \\
\textit{$\mathtt{\ }$\hspace{0.4cm} $\alpha F(s) + \beta G(s) $     }
 \end{array}$}  }  &

   \multicolumn{1}{l}{{$\begin{array} {lcl} \textup{\texttt{\hspace{0.0cm}$\vdash$ $\forall$ f g s a b.  }} \\
   \textup{\texttt{$\mathtt{\ }$ laplace\_exists f s $\wedge$ laplace\_exists g s  \hspace{-0.5cm}}} \\
\textup{\texttt{$\mathtt{\ }$\hspace{0.2cm} $\Rightarrow$  laplace\_transform ($\lambda$t. a $\ast$ f t + b $\ast$ g t) s = \hspace{-0.5cm} }} \\
\textup{\texttt{$\mathtt{\ }$\hspace{1.20cm} a $\ast$ laplace\_transform f s +  \hspace{-0.5cm} }} \\
\textup{\texttt{$\mathtt{\ }$\hspace{2.00cm} b $\ast$ laplace\_transform g s \hspace{-0.5cm} }}
 \end{array}$}}    \\ \hline




   \multicolumn{1}{l|}{ {$\begin{array} {lcl} \textbf{Frequency Shifting} \\
   \hspace{0.0cm} \textit{$ \mathcal{L} [ e^{s_0 t} f(t)] = $ } \\
\textit{$\mathtt{\ }$\hspace{0.4cm} $ F(s - s_0) $     }
 \end{array}$}  }  &


   \multicolumn{1}{l}{{$\begin{array} {lcl} \hspace{0.00cm} \textup{\texttt{$\vdash$ $\forall$ f s s0. laplace\_exists f s      \hspace{-0.5cm} }} \\
\textup{\texttt{$\mathtt{\ }$\hspace{-0.4cm} $\Rightarrow$ laplace\_transform  \hspace{-0.5cm} }} \\
\textup{\texttt{$\mathtt{\ }$\hspace{0.1cm} ($\lambda$t. cexp (s0 $\ast$ Cx (drop t)) $\ast$ f t) s = \hspace{-0.5cm} }} \\
\textup{\texttt{$\mathtt{\ }$\hspace{2.0cm} laplace\_transform f (s - s0) \hspace{-0.5cm} }}
 \end{array}$}}    \\ \hline




     \multicolumn{1}{l|}{{$\begin{array} {lcl} \textbf{First-order Differ-} \\
     \textbf{entiation in Time} \\
     \textbf{Domain} \\
  \hspace{0.0cm}  \mathcal{L} \left[\dfrac{d}{dt}f(t) \right] = \\
  \hspace{0.1cm}   s F(s) - f(0)
     \end{array}$}}    &

    \multicolumn{1}{l}{{$\begin{array} {lcl} \textup{\texttt{ \hspace{0.00cm} $\vdash$ $\forall$ f s. laplace\_exists f s $\wedge$  \hspace{-0.5cm} }} \\
\textup{\texttt{$\mathtt{\ }$\hspace{-0.2cm} ($\forall$t. f differentiable at t) $\wedge$  \hspace{-0.5cm} }}  \\
\textup{\texttt{$\mathtt{\ }$\hspace{-0.2cm} laplace\_exists ($\lambda$t. vector\_derivative f (at t)) s \hspace{-0.5cm}  }}  \\
\textup{\texttt{$\mathtt{\ }$\hspace{-0.2cm} $\Rightarrow$ laplace\_transform  \hspace{-0.5cm} }}  \\
\textup{\texttt{$\mathtt{\ }$\hspace{0.2cm} ($\lambda$t. vector\_derivative f (at t)) s = \hspace{-0.5cm}  }}  \\
\textup{\texttt{$\mathtt{\ }$\hspace{0.8cm} s $\ast$ laplace\_transform f s - f (lift (\&0)) \hspace{-0.5cm} }}
 \end{array}$}}    \\ \hline




    \multicolumn{1}{l|}{{$\begin{array} {lcl} \textbf{Higher-order Diffe-} \\
     \textbf{rentiation in Time} \\
     \textbf{Domain} \\
  \hspace{0.0cm}  \mathcal{L} [\dfrac{d^n}{{dt}^n}f(t)] = s^n F(s) \\
  \hspace{-0.2cm}  - \sum_{k = 1}^{n}{ s^{k - 1} \dfrac{d^{n - k} f (0)}{{dx}^{n - k}} }
     \end{array}$}}    &

    \multicolumn{1}{l}{{$\begin{array} {lcl} \textup{\texttt{\hspace{0.00cm} $\vdash$ $\forall$ f s n. laplace\_exists\_higher\_deriv n f s $\wedge$ \hspace{-0.5cm} }} \\
\textup{\texttt{$\mathtt{\ }$\hspace{0.1cm} ($\forall$t. differentiable\_higher\_derivative n f t) \hspace{-0.5cm}  }} \\
\textup{\texttt{$\mathtt{\ }$\hspace{0.1cm} $\Rightarrow$ laplace\_transform  \hspace{-0.5cm} }} \\
\textup{\texttt{$\mathtt{\ }$\hspace{1.0cm} ($\lambda$t. higher\_vector\_derivative n f t) s = \hspace{-0.5cm} }} \\
\textup{\texttt{$\mathtt{\ }$\hspace{0.0cm} s pow n $\ast$ laplace\_transform f s -  \hspace{-0.5cm} }} \\
\textup{\texttt{$\mathtt{\ }$\hspace{0.2cm} vsum (1..n) ($\lambda$x. s pow (x - 1) $\ast$ \hspace{-0.5cm} }} \\
\textup{\texttt{$\mathtt{\ }$\hspace{0.4cm} higher\_vector\_derivative (n - x) f (lift (\&0)))  \hspace{-0.5cm} }}
 \end{array}$}}    \\ \hline




    \multicolumn{1}{l|}{{$\begin{array} {lcl} \textbf{Integration in} \\
     \textbf{Time Domain} \\
  \mathcal{L} \left[ \int_{0}^{t}{f (\tau) d\tau} \right] = \dfrac{1}{s} F(s)
     \end{array}$}}    &


    \multicolumn{1}{l}{{$\begin{array} {lcl} \textup{\texttt{\hspace{0.0cm}$\vdash$ $\forall$ f s. \&0 < Re s $\wedge$ laplace\_exists f s $\wedge$   }}  \\
\textup{\texttt{$\mathtt{\ }$\hspace{0.0cm} laplace\_exists   }}  \\
\textup{\texttt{$\mathtt{\ }$\hspace{0.2cm} ($\lambda$x. integral (interval [lift (\&0),x]) f) s $\wedge$ \hspace{-0.5cm}  }}  \\
\textup{\texttt{$\mathtt{\ }$\hspace{0.0cm} ($\forall$x. f continuous\_on interval [lift (\&0),x]) \hspace{-0.5cm}  }}  \\
\textup{\texttt{$\mathtt{\ }$\hspace{-0.2cm} $\Rightarrow$ laplace\_transform  \hspace{-0.5cm} }}  \\
\textup{\texttt{$\mathtt{\ }$\hspace{0.4cm} ($\lambda$x. integral (interval [lift (\&0),x]) f) s =  \hspace{-0.5cm} }}  \\
\textup{\texttt{$\mathtt{\ }$\hspace{2.0cm}  $\mathtt{\dfrac{Cx (\&1)}{s}}$ $\ast$ laplace\_transform f s \hspace{-0.5cm}  }}
\end{array}$}}    \\ \hline




   \multicolumn{1}{l|}{ {$\begin{array} {lcl} \textbf{Time Shifting} \\
    \mathcal{L} \left[ f(t - t_0) u(t - t_0) \right] =  \\
   \textit{$\mathtt{\ }$\hspace{0.4cm} $  e^{-t_0 s} F(s) $    } \\

 \end{array}$}  }  &

    \multicolumn{1}{l}{{$\begin{array} {lcl} \textup{\texttt{\hspace{0.0cm} $\vdash$ $\forall$ f s t0. \&0 < drop t0 $\wedge$ laplace\_exists f s \hspace{-0.5cm}  }} \\
\textup{\texttt{$\mathtt{\ }$\hspace{0.1cm} $\Rightarrow$ laplace\_transform (shifted\_fun f t0) s = \hspace{-0.5cm}  }} \\
\textup{\texttt{$\mathtt{\ }$\hspace{0.6cm} cexp (--(s $\ast$ Cx (drop t0))) $\ast$ \hspace{-0.5cm} }} \\
\textup{\texttt{$\mathtt{\ }$\hspace{3.0cm} laplace\_transform f s  \hspace{-0.5cm}  }}
 \end{array}$}}    \\ \hline





    \multicolumn{1}{l|}{ {$\begin{array} {lcl} \textbf{Time Scaling} \\
     \mathcal{L} \left[ f(c t) \right] = \dfrac{1}{c} F\left(\dfrac{s}{c}\right),  \\
   \textit{$\mathtt{\ }$\hspace{0.4cm} $ \ \ 0 < c $  } \\
 \end{array}$}  }  &

    \multicolumn{1}{l}{{$\begin{array} {lcl} \textup{\texttt{$\vdash$   $\forall$ f s c. \&0 < c $\wedge$ laplace\_exists f s $\wedge$    \hspace{-0.5cm}  }} \\
    \textup{\texttt{$\mathtt{\ }$\hspace{0.3cm} laplace\_exists f$\left(\mathtt{\dfrac{s}{Cx\ c}}\right)$   \hspace{-0.5cm} }} \\
\textup{\texttt{$\mathtt{\ }$\hspace{0.0cm} $\Rightarrow$  laplace\_transform ($\lambda$t. f(c \% t)) s =  \hspace{-0.5cm} }} \\
\textup{\texttt{$\mathtt{\ }$\hspace{2.0cm} $\mathtt{\dfrac{Cx (\&1)}{Cx\ c}}$$\ast$laplace\_transform f$\left(\mathtt{\dfrac{s}{Cx\ c}}\right)$  \hspace{-0.5cm} }}
 \end{array}$}}    \\ \hline




   \multicolumn{1}{l|}{ {$\begin{array} {lcl} \textbf{Cosine Based} \\
     \textbf{Modulation} \\
     \mathcal{L} \left[ f(t) cos(\omega_0t) \right] =  \\
   \textit{$\mathtt{\ }$\hspace{-0.1cm} $  \dfrac{F(s - j\omega_0)}{2} \ + $     } \\
   \textit{$\mathtt{\ }$\hspace{1.0cm} $ \dfrac{F(s + j\omega_0)}{2} $    }
 \end{array}$}  }  &

    \multicolumn{1}{l}{{$\begin{array} {lcl} \textup{\texttt{\hspace{0.0cm}$\vdash$   $\forall$ f s w0. laplace\_exists f s }} \\
\textup{\texttt{$\mathtt{\ }$\hspace{-0.25cm}  $\Rightarrow$  laplace\_transform     }}  \\
\textup{\texttt{$\mathtt{\ }$\hspace{0.5cm}   ($\lambda$t. ccos (Cx w0 $\ast$ Cx (drop t)) $\ast$ f t) s =  \hspace{-0.5cm}  }}  \\
\textup{\texttt{$\mathtt{\ }$\hspace{1.0cm}  $\mathtt{\dfrac{laplace\_transform\ f\ (s - ii \ast Cx\ w0)}{Cx (\&2)}\ +}$ \hspace{-0.4cm}  \hspace{-0.5cm} }} \\
\textup{\texttt{$\mathtt{\ }$\hspace{2.0cm}  $\mathtt{\dfrac{laplace\_transform\ f\ (s + ii \ast Cx\ w0)}{Cx (\&2)}}$ \hspace{-0.4cm}  \hspace{-0.5cm} }}
 \end{array}$}}    \\ \hline


   \multicolumn{1}{l|}{ {$\begin{array} {lcl} \textbf{Sine Based} \\
     \textbf{Modulation} \\
     \mathcal{L} \left[ f(t) cos(\omega_0t) \right] =  \\
   \textit{$\mathtt{\ }$\hspace{-0.1cm} $  \dfrac{F(s - j\omega_0)}{2j} \ - $     } \\
   \textit{$\mathtt{\ }$\hspace{1.0cm} $ \dfrac{F(s + j\omega_0)}{2j} $    }
 \end{array}$}  }  &

    \multicolumn{1}{l}{{$\begin{array} {lcl} \textup{\texttt{ \hspace{0.0cm}$\vdash$   $\forall$ f s w0. laplace\_exists f s $\Rightarrow$ }} \\
\textup{\texttt{$\mathtt{\ }$\hspace{-0.25cm}   laplace\_transform     }}  \\
\textup{\texttt{$\mathtt{\ }$\hspace{0.5cm}   ($\lambda$t. csin (Cx w0 $\ast$ Cx (drop t)) $\ast$ f t) s =  \hspace{-0.5cm}  }}  \\

\textup{\texttt{$\mathtt{\ }$\hspace{1.0cm}  $\mathtt{\dfrac{laplace\_transform\ f\ (s - ii \ast Cx\ w0)}{Cx (\&2) \ast ii}} \ -$ \hspace{-0.4cm}  \hspace{-0.5cm} }}  \\
\textup{\texttt{$\mathtt{\ }$\hspace{2.0cm}  $\mathtt{\dfrac{laplace\_transform\ f\ (s + ii \ast Cx\ w0)}{Cx (\&2) \ast ii}}$ \hspace{-0.4cm}  \hspace{-0.5cm} }}
 \end{array}$}}    \\ \hline




   \multicolumn{1}{l|}{ {$\begin{array} {lcl} \textbf{$n$-order Differ-} \\
     \textbf{ential Equation} \\
     \mathcal{L} \Big( \sum _{k = 0}^{n} {{\alpha}_k \dfrac{d^ky}{{dt}^k}} \Big) =    \\
\textit{$\mathtt{\ }$\hspace{0.4cm} $F(s) \ \sum _{k = 0}^{n} {{\alpha}_k s^k} $    } \\
\textit{$\mathtt{\ }$\hspace{-0.2cm} $- \sum _{k = 0}^{n} {\sum_{i = 1}^{k}}$    } \\
\textit{$\mathtt{\ }$\hspace{0.5cm} ${s^{i - 1} \dfrac{d^{k - i}f(0)}{{dt}^{k - i}} }$    }
 \end{array}$}  }  &

    \multicolumn{1}{l}{{$\begin{array} {lcl} \textup{\texttt{$\vdash$ $\forall$ f lst s n. laplace\_exists\_higher\_deriv n f s $\wedge$  }} \\
\textup{\texttt{$\mathtt{\ }$\hspace{0.1cm} ($\forall$t. differentiable\_higher\_derivative n f t)  }} \\
\textup{\texttt{$\mathtt{\ }$\hspace{0.1cm} $\Rightarrow$ laplace\_transform    }} \\
\textup{\texttt{$\mathtt{\ }$\hspace{0.9cm} ($\lambda$t. diff\_eq\_n\_order n lst f t) s =   }} \\

\textup{\texttt{$\mathtt{\ }$\hspace{0.1cm} laplace\_transform f s $\ast$  \hspace{-0.5cm}  }} \\
\textup{\texttt{$\mathtt{\ }$\hspace{0.5cm} vsum (0..n) ($\lambda$k. EL k lst $\ast$ s pow k) \hspace{-0.5cm}  }} \\

\textup{\texttt{$\mathtt{\ }$\hspace{0.1cm} - vsum (0..n) ($\lambda$k. EL k lst $\ast$ \hspace{-0.5cm}  }} \\
\textup{\texttt{$\mathtt{\ }$\hspace{0.5cm} vsum (1..k) ($\lambda$i. s pow (i - 1)  \hspace{-0.5cm} }} \\
\textup{\texttt{$\mathtt{\ }$\hspace{-0.1cm} $\ast$ higher\_vector\_derivative (k - i) f (lift (\&0))))  \hspace{-0.5cm} }}
 \end{array}$}}    \\ \hline


    \end{longtable}

\end{footnotesize}


\noindent  place of all the derivatives up to the $n^{th}$ order of the function \texttt{f} exist.  Similarly, the predicate \texttt{differentiable\_higher\_derivative}  provides the differentiability of the function \texttt{f} and its higher derivatives up to the $n^{th}$ order. The verification of these properties not only ensures the correctness of our definitions but also plays a vital role in minimizing the user effort in reasoning about Laplace transform based analysis of systems, as will be depicted in Sections~\ref{SEC:cs_properties} and~\ref{SEC:unmanned_vehicle} of this paper.

The generalized linear differential equation describes the input-output relationship for a generic $n$-order linear control system~\cite{nise2007control}:

\small

\begin{equation}\label{EQ:diff_eqn_nth_order_LTI_sys}
 \sum _{k = 0}^{n} {{\alpha}_k \dfrac{d^k}{{dt}^k} y(t)} = \sum _{k = 0}^{m} {{\beta}_k \dfrac{d^k}{{dt}^k} x(t)}, \ \ \ \  m \leq n
\end{equation}

\normalsize

\noindent where $y(t)$ is the output and $x(t)$ is the input to the system. The constants $\alpha_k$ and
$\beta_k$ are the coefficients of the output and input differentials with order $k$, respectively. The greatest index $n$ of
the non-zero coefficient $\alpha_n$ determines the order of the underlying system.
The corresponding transfer function is obtained by setting the initial conditions equal to zero~\cite{nise2007control}:


\small

\begin{equation}\label{EQ:transfer_fun_nth_order_LTI_sys}
\dfrac{Y(s)}{X(s)} = \dfrac{\sum_{k = 0}^{m} {\beta_k s^k}}{\sum_{k = 0}^{n} {\alpha_k s^k}}
\end{equation}

\normalsize

We verified the transfer function, given in Equation~\ref{EQ:transfer_fun_nth_order_LTI_sys}, for the generic n-order linear control system as the following HOL Light theorem.

	\begin{flushleft}
		\begin{theorem}
			\label{THM:transfer_fun_n_order_sys}
{\footnotesize
				\textup{\texttt{$\vdash$ $\forall$ y x m n inlst outlst s.   \\
						$\mathtt{\ }$\hspace{-0.1cm} ($\forall$t. differentiable\_higher\_deriv m n x y t) $\wedge$  \\
						$\mathtt{\ }$\hspace{-0.1cm} laplace\_exists\_of\_higher\_deriv m n x y s $\wedge$ zero\_init\_conditions m n x y $\wedge$  \\
						$\mathtt{\ }$\hspace{-0.1cm} diff\_eq\_n\_order\_sys m n inlst outlst y x $\wedge$  \\
						$\mathtt{\ }$\hspace{-0.1cm} $\sim$(laplace\_transform x s = Cx (\&0)) $\wedge$  \\
						$\mathtt{\ }$\hspace{-0.1cm} $\sim$(vsum (0..n) ($\lambda$t. EL t outlst $\ast$ s pow t) = Cx (\&0))  \\
						$\mathtt{\ }$\hspace{1.0cm} $\Rightarrow$ $\mathtt{\dfrac{laplace\_transform\ y\ s}{laplace\_transform\ x\ s}}$ =       $\mathtt{\dfrac{vsum\ (0..m)\ (\lambda t.\ EL\ t\ inlst\ \ast\ s\ pow\ t)}{vsum\ (0..n)\ (\lambda t.\ EL\ t\ outlst\ \ast\ s\ pow\ t)}}$
			}}}
		\end{theorem}
	\end{flushleft}

\noindent The first assumption ensures that the functions $\texttt{y}$ and $\texttt{x}$ are differentiable up to the $n^{th}$ and $m^{th}$
order, respectively. The next assumption represents the Laplace transform existence condition up to the $n^{th}$ order derivative
of function $\texttt{y}$ and $m^{th}$ order derivative of the function $\texttt{x}$. The next assumption models the zero initial
conditions for both of the functions $\texttt{y}$ and $\texttt{x}$, respectively. The next assumption represents the
formalization of Equation \ref{EQ:diff_eqn_nth_order_LTI_sys} and the last two assumptions provide the conditions for the design of a reliable linear control system.
Finally, the conclusion of the above theorem represents the transfer function given by Equation \ref{EQ:transfer_fun_nth_order_LTI_sys}.
The verification of this theorem is very useful as it allows to automate the verification of the transfer function of any linear control system as described in Sections~\ref{SEC:cs_properties} and~\ref{SEC:unmanned_vehicle} of the paper.
The formalization, described in this section, took around $2000$ lines of HOL Light code~\cite{adnan2016facstp_wp} and around $130$ man-hours.

\section{Formalization of Linear Control Systems Foundations} \label{SEC:cs_properties}

\begin{figure}
\centering
\captionsetup[subfigure]{oneside,margin={0.0cm,0cm}}
\subfloat[ ]{
\includegraphics[width=0.34\linewidth, trim={9 9 9 9},clip]{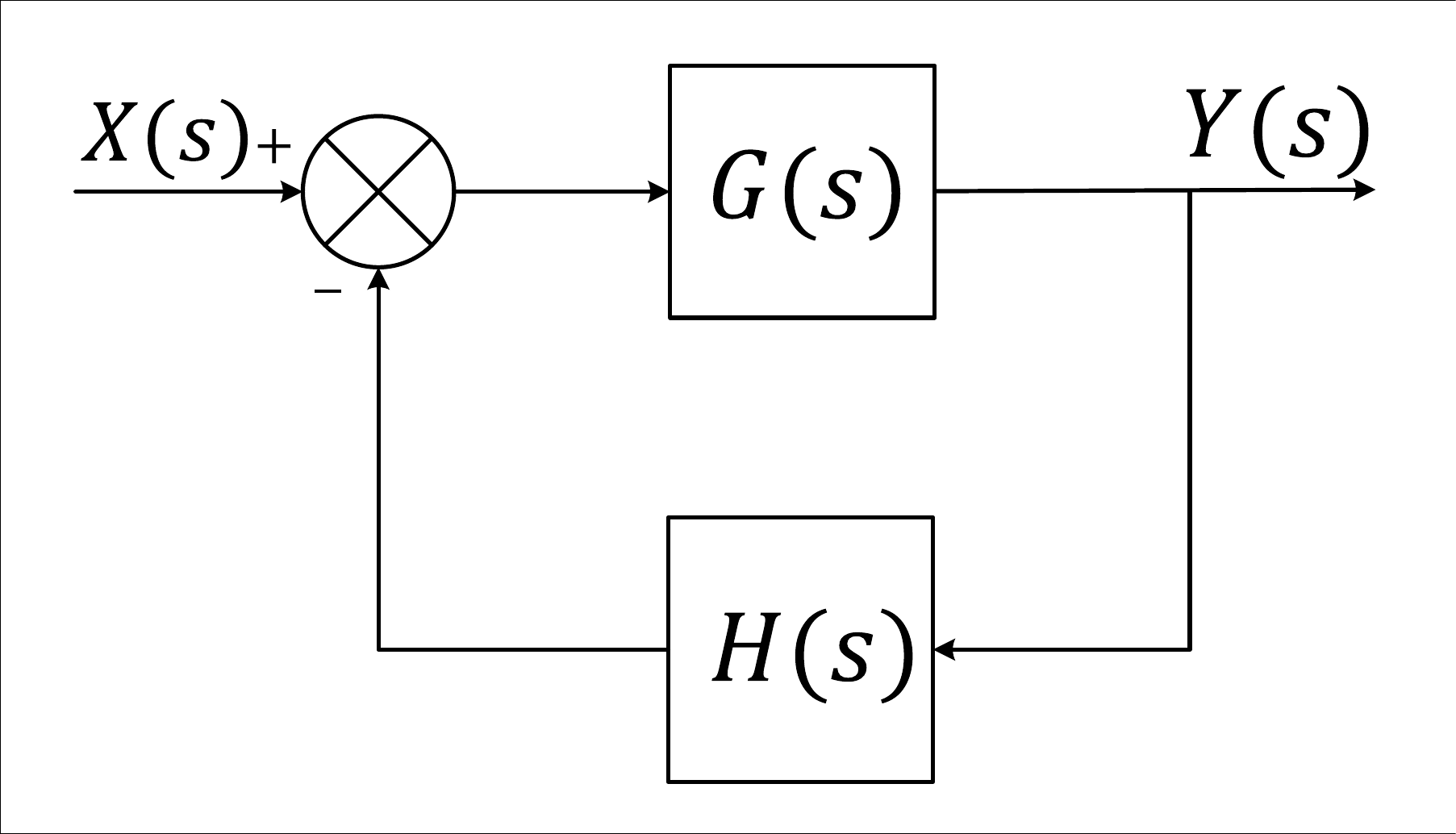}%
\label{FIG:closed_loop_cs}
}
\captionsetup[subfigure]{oneside,margin={0.0cm,0cm}}
\subfloat[ ]{
\includegraphics[width=0.42\linewidth, trim={9 9 9 9},clip]{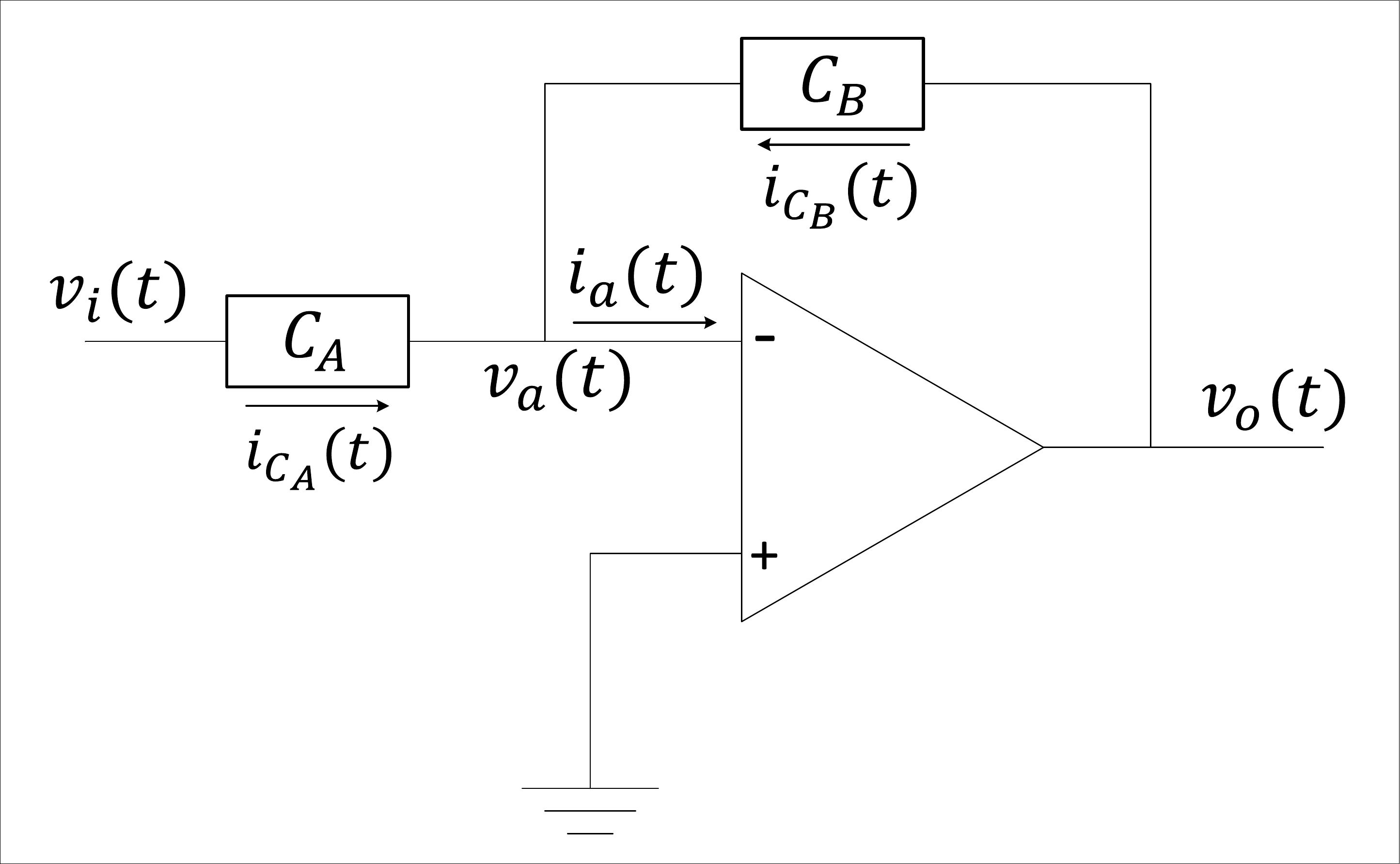}%
\label{FIG:generic_active_realization}
}
\captionsetup[subfigure]{oneside,margin={0.0cm,0cm}}
\subfloat[ ]{
\includegraphics[width=0.16\linewidth, trim={9 9 9 9},clip]{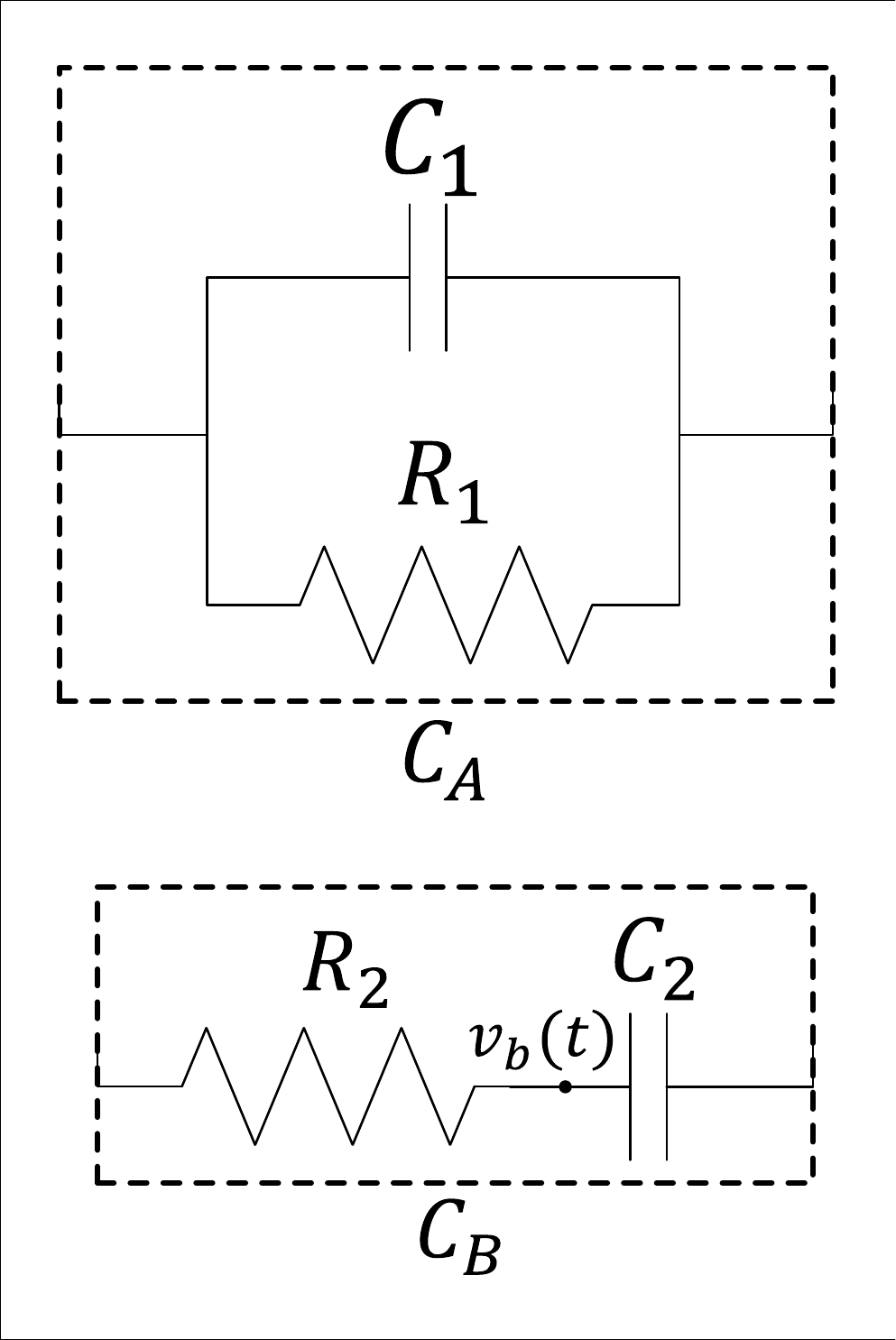}%
\label{FIG:config_pid_controller}
}
\captionsetup[subfigure]{oneside,margin={0.0cm,0cm}}
\subfloat[ ]{
\includegraphics[width=0.33\linewidth, trim={9 9 9 9},clip]{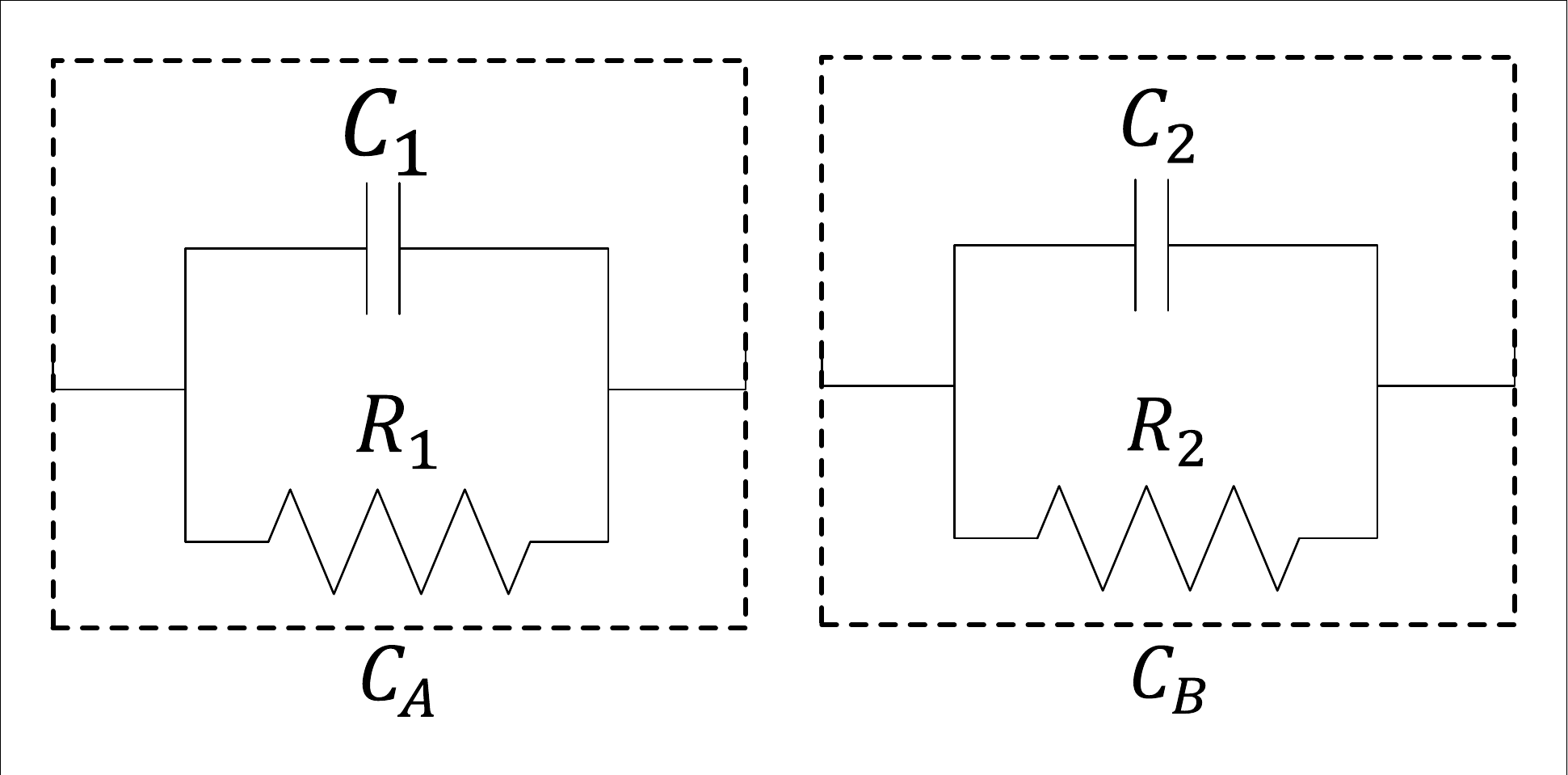}%
\label{FIG:config_compensator}
}
\captionsetup[subfigure]{oneside,margin={0.0cm,0cm}}
\subfloat[ ]{
\includegraphics[width=0.28\linewidth, trim={9 9 9 9},clip]{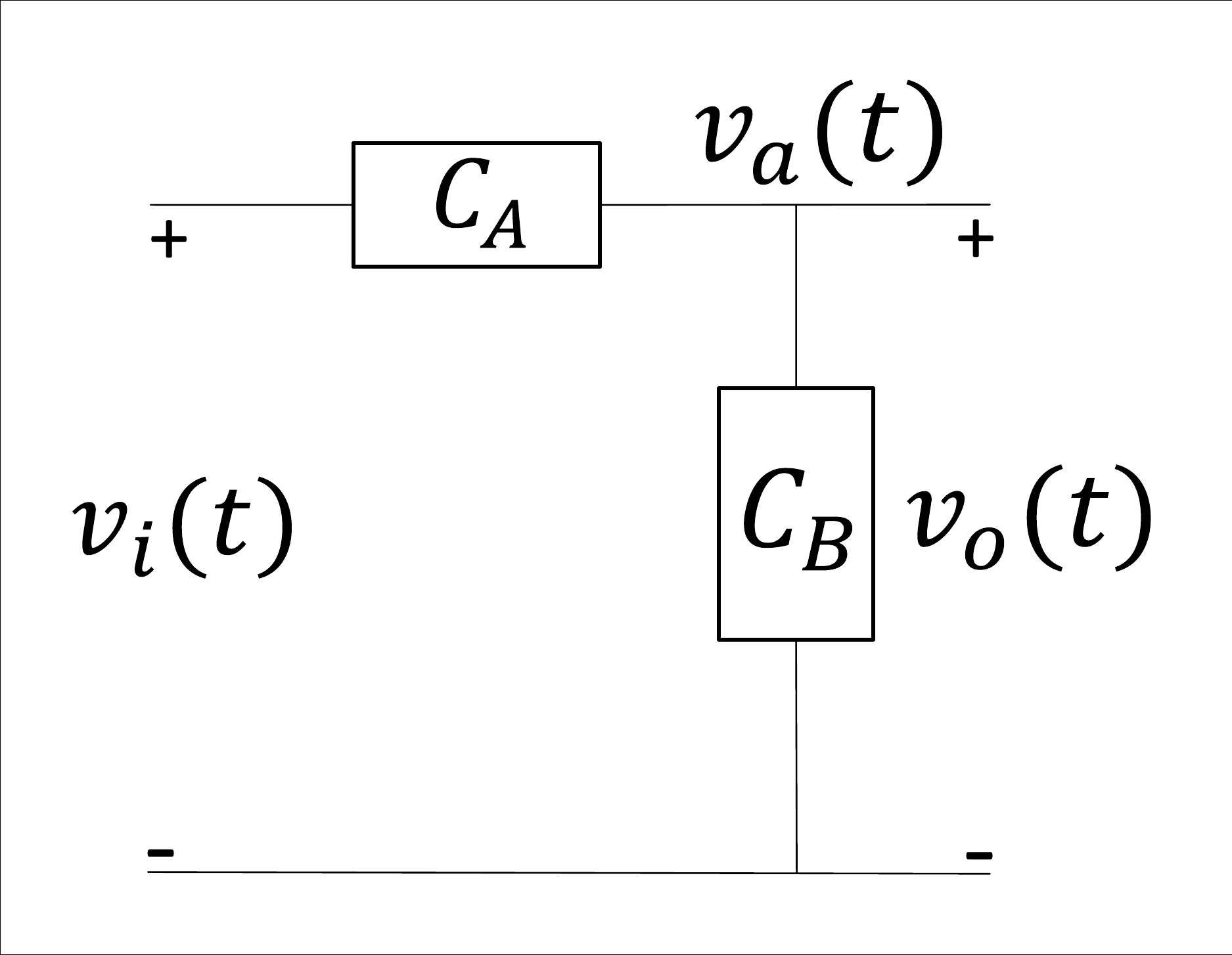}%
\label{FIG:generic_passive_realization}
}
\captionsetup[subfigure]{oneside,margin={0.0cm,0cm}}
\subfloat[ ]{
\includegraphics[width=0.28\linewidth, trim={9 9 9 9},clip]{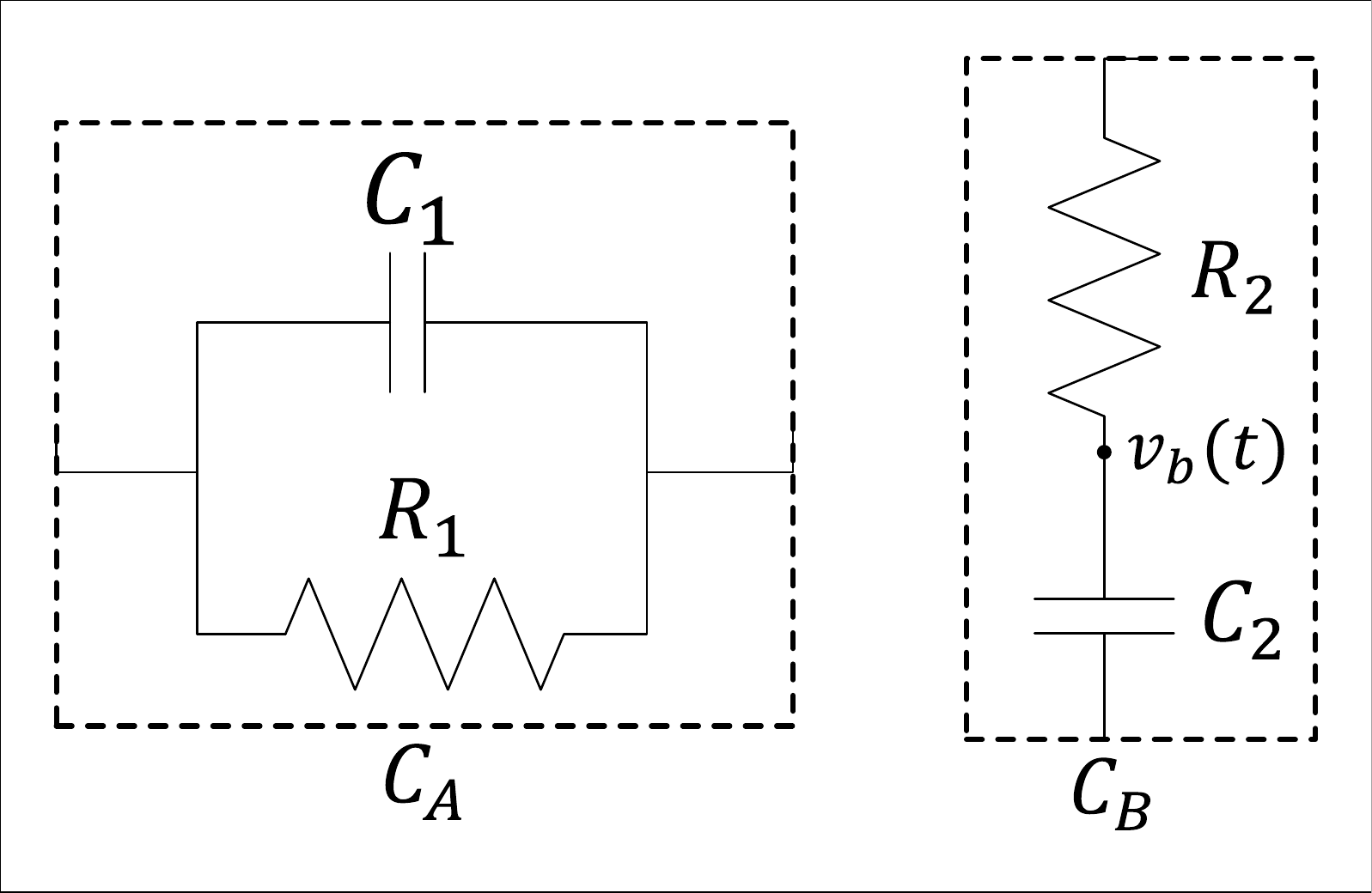}%
\label{FIG:config_lead_lag_compensator}
}
\caption{Control Systems Foundations (a) Closed Loop Control System (b) Generic Active Realization of Controller (c) PID Configuration (d) Lag/lead Compensator Configuration (e) Generic Passive Realization of Compensator (f) Lag-lead Compensator Configuration}
\label{FIG:controller}
\end{figure}

A general closed-loop control system is depicted in Fig.~\ref{FIG:closed_loop_cs}. Here, $X(s)$ and $Y(s)$ represent the Laplace transforms of the time domain input $x(t)$ and the output $y(t)$, respectively. $G(s)$ and $H(s)$ represent the forward path and the feedback path transfer functions, respectively. Similarly, $G(s) H(s)$ is the open loop transfer function of the system and
$Y(s)/X(s)$ is the closed loop transfer function~\cite{ghosh2010control}.
Table~\ref{TAB:properties of_control_systems} presents the formalization of the frequency response, phase margin and gain margin of this control system. These properties are used to study the dynamics of a linear control system in the frequency domain and to perform its stability analysis.

The frequency response is used to analyze the dynamics of the system by studying the impact of different frequency components on the intended behaviour of the given linear control system.
We also formally verified the frequency response of a generic $n$-order system based on assumptions that are very similar to the ones used for Theorem~\ref{THM:transfer_fun_n_order_sys}.


\begin{footnotesize}
    \begin{longtable}{|p{2cm}|p{3cm}|p{7cm}|p{3cm}|}
\caption{Properties of Linear Control Systems}
\label{TAB:properties of_control_systems}
\endfirsthead
\endhead
    \hline
    \hline
    \multicolumn{1}{l|}{Property}   &
    \multicolumn{1}{l}{\hspace{0.0cm} Formalized Form}

     \\ \hline \hline




    \multicolumn{1}{l|}{ {$\begin{array} {lcl} \textbf{Frequency Response} \\
     M (j\omega) = M (s)|_{(j\omega)} =  \\
\textit{$\mathtt{\ }$\hspace{0.4cm} $\dfrac{Y(s)}{X(s)}\Bigg{|}_{(j\omega)} = \dfrac{Y(j\omega)}{X(j\omega)}$     }
 \end{array}$}   }  &

   \multicolumn{1}{l}{{$\begin{array} {lcl} \textup{\texttt{\hspace{0.0cm}$\vdash$ $\forall$ y x w. frequency\_response x y w =   }} \\
\textup{\texttt{$\mathtt{\ }$\hspace{0.2cm}  $\mathtt{\dfrac{laplace\_transform\ y\ (ii\ \ast\ Cx\ w)}{laplace\_transform\ x\ (ii\ \ast\ Cx\ w)}}$  }}
 \end{array}$}}    \\ \hline




    \multicolumn{1}{l|}{ {$\begin{array} {lcl} \textbf{Frequency Response} \\
     \textbf{of an $n$-order} \\
     \textbf{System} \\
     \dfrac{Y(j\omega)}{X(j\omega)} = \dfrac{\sum_{k = 0}^{m} {\beta_k (j\omega)^k}}{\sum_{k = 0}^{n} {\alpha_k (j\omega)^k}}
 \end{array}$}  }  &

   \multicolumn{1}{l}{ $\begin{array} {lcl} \textup{\texttt{\hspace{0.0cm}$\vdash$ $\forall$ y x m n inlst outlst s.    }} \\
\textup{\texttt{$\mathtt{\ }$\hspace{0.2cm} ($\forall$t. differentiable\_higher\_deriv m n x y t) $\wedge$     }} \\
\textup{\texttt{$\mathtt{\ }$\hspace{0.2cm} laplace\_exists\_of\_higher\_deriv m n x y w $\wedge$     }} \\
\textup{\texttt{$\mathtt{\ }$\hspace{0.2cm} zero\_init\_conditions m n x y $\wedge$     }} \\
\textup{\texttt{$\mathtt{\ }$\hspace{0.2cm} diff\_eq\_n\_order\_sys m n inlst outlst y x $\wedge$   }} \\
\textup{\texttt{$\mathtt{\ }$\hspace{0.2cm} non\_zero\_denom\_cond n x w outlst $\Rightarrow$  }} \\
\textup{\texttt{$\mathtt{\ }$\hspace{-0.2cm} frequency\_response x y w =    }} \\
\textup{\texttt{$\mathtt{\ }$\hspace{0.0cm} $\mathtt{\dfrac{vsum\ (0..m)\ (\lambda t.\ EL\ t\ inlst\ \ast\ (ii \ast Cx\ w)\ pow\ t)}{vsum\ (0..n)\ (\lambda t.\ EL\ t\ outlst\ \ast\ (ii \ast Cx\ w)\ pow\ t)}}$  }}
 \end{array}$ }    \\ \hline




    \multicolumn{1}{l|}{ {$\begin{array} {lcl} \textbf{Phase Margin} \\ {[\angle G(j\omega)H(j\omega)]_{\omega = \omega_{gc}} }  \\
\textit{$\mathtt{\ }$\hspace{0.4cm} $+\ 180^o$     }
 \end{array}$}   }  &


   \multicolumn{1}{l}{{$\begin{array} {lcl} \textup{\texttt{\hspace{0.0cm}$\vdash$ $\forall$ g h wgc.  phase\_margin g h wgc =  }} \\
\textup{\texttt{$\mathtt{\ }$\hspace{0.2cm} pi + Arg (g (ii $\ast$ Cx wgc) $\ast$ h (ii $\ast$ Cx wgc))  }}
 \end{array}$}}    \\ \hline




    \multicolumn{1}{l|}{ {$\begin{array} {lcl} \textbf{Gain Margin} \\ \Big[20 \textit{log}_{10}\Big|G(j\omega)  \\
\textit{$\mathtt{\ }$\hspace{0.6cm} $H(j\omega) \Big|_{\omega = \omega_{pc}}\Big]dB $   }
 \end{array}$}   }  &


   \multicolumn{1}{l}{ {$\begin{array} {lcl} \textup{\texttt{\hspace{0.0cm}$\vdash$ $\forall$ g h wpc.  gain\_margin\_db g h wpc = \&20 $\ast$  }} \\
\textup{\texttt{$\mathtt{\ }$\hspace{-0.1cm} $\mathtt{\dfrac{log\ (norm\ (g\ (ii\ \ast\ Cx\ wpc)\ \ast\ h\ (ii\ \ast\ Cx\ wpc)))}{log\ (\&10)}}$  }}
 \end{array}$}  }    \\ \hline


    \end{longtable}

\end{footnotesize}


Phase margin and gain margin provide useful information about controlling the stability of the system~\cite{ghosh2010control}. Phase margin represents $180^o$ shifted phase angle of the open loop transfer function evaluated at the gain crossover frequency ($\omega_{gc}$), which is the frequency at which the magnitude of the open loop transfer function is equal to $0$ dB.
The gain margin represents the magnitude of the open loop transfer function evaluated at the phase crossover frequency ($\omega_{pc}$), which is the frequency at which the resultant phase curve of the open loop gain has a phase of $180^o$. In our formal definitions of these notions, the function \texttt{Arg(z)} represents the argument of a complex number \texttt{z}.

\vspace{-1.0cm}

\vspace{-0.50cm}

The controllers form the most vital part of any control system as they are mainly responsible for the correct operation of every component of the underlying system. Controllers are modeled using their active realizations based on an electrical circuit, which comprises of an inverting operational amplifier (op-amp) with unity gain, and two components, i.e., $C_A$ and $C_B$, which are shown as rectangular boxes in
Fig.~\ref{FIG:generic_active_realization}.
The boxes $C_A$ and $C_B$ contain different configurations of the passive
components, i.e., resistors and capacitors~\cite{ogata1970modern}. By making an appropriate choice of these passive components,
we obtain various controllers, such as P, I, D, PI, PD, PID~\cite{nise2007control}. For the analysis of these controllers, we first need to formalize them in higher-order logic. This step requires a formal library of analog components~\cite{taqdees2017tflac,adnan2016facstp_wp}, describing the voltage-current relationships of resistor, capacitors and inductors, and the KCL and KVL, which model the currents and voltages in an electrical circuit.

The PID controller, depicted in Fig.~\ref{FIG:config_pid_controller}, can be formalized as follows:

\begin{flushleft}
\begin{definition}
\label{DEF:pid_implem_model}
{\footnotesize
				\textup{\texttt{$\vdash$ $\forall$ C1 R1 Vi R2 C2 Vo Vb Va.   \\
						$\mathtt{\ }$\hspace{0.0cm} pid\_controller\_implem Vi Vo Va Vb C1 C2 R1 R2 $\Leftrightarrow$  \\
						$\mathtt{\ }$\hspace{0.0cm} ($\forall$t. \&0 < drop t $\Rightarrow$ kcl [$\lambda$t. capacitor\_current C1 ($\lambda$t. Vi t - Va t) t;   \\
						$\mathtt{\ }$\hspace{3.50cm} $\lambda$t. resistor\_current R1 ($\lambda$t. Vi t - Va t) t; \\
						$\mathtt{\ }$\hspace{3.50cm} $\lambda$t. resistor\_current R2 ($\lambda$t. Vb t - Va t) t] t $\wedge$ \\
						$\mathtt{\ }$\hspace{0.0cm} ($\forall$t. \&0 < drop t $\Rightarrow$ kcl [$\lambda$t. resistor\_current R2 ($\lambda$t. Va t - Vb t) t;   \\
						$\mathtt{\ }$\hspace{3.50cm} $\lambda$t. capacitor\_current C2 ($\lambda$t. Vo t - Vb t) t] t $\wedge$ \\
						$\mathtt{\ }$\hspace{0.0cm} ($\forall$t. \&0 < drop t $\Rightarrow$ Va t = Cx (\&0))
   }}}
\end{definition}
\end{flushleft}

\noindent where $\texttt{Vi}$ and $\texttt{Vo}$ are the input and the output voltages, respectively, having data type $\mathtt{\mathds{R}^1 \rightarrow \mathds{C}}$, and $\texttt{Va}$ and $\texttt{Vb}$ are the voltages at nodes $a$ and $b$, respectively. The functions $\texttt{resistor\_current}$ and $\texttt{capcitor\_current}$ are the currents across the resistor and capacitor, respectively. The function $\texttt{kcl}$ accepts a list of currents across the components of the circuit and a time variable $\texttt{t}$ and returns the predicate that guarantees that the sum of all the currents leaving a particular node at time $\texttt{t}$ is zero. The first conjunct of the above definition represents the application of KCL across node $a$. Similarly, the second conjunct models the KCL at node $b$, whereas the last conjunct provides the voltage across the non-inverting input of the op-amp using the virtual ground condition, as shown in Fig.~\ref{FIG:generic_active_realization}. We also develop a simplification tactic $\texttt{KCL\_SIMP\_TAC}$, which simplifies the implementations of the PID controller as well as other controllers and compensators. The details can be found in~\cite{adnan2016facstp_wp}.

Next, we model the dynamical behaviour of the PID controller using the $n$-order differential equation:

\begin{flushleft}
\begin{definition}
\label{DEF:pid_behav_model}
{\footnotesize
\textup{\texttt{$\vdash$ $\forall$ R1 R2 C1 C2. inlst\_pid\_contr R1 R2 C1 C2 =  }}} \\
$\mathtt{\ }$\hspace{0.2cm}  \small\textup{\texttt{[--Cx (\&1); --Cx (R2 $\ast$ C2 + R1 $\ast$ C1); --Cx (R1 $\ast$ R2 $\ast$ C1 $\ast$ C2)]}} \\
{\footnotesize
\textup{\texttt{$\vdash$ $\forall$ R1 C2. outlst\_pid\_contr R1 C2 = [Cx (\&0); Cx (R1 $\ast$ C2)]  }}}\\
{\footnotesize
\textup{\texttt{$\vdash$ $\forall$ Vo R1 R2 C1 C2 Vi t. pid\_controller\_behav\_spec R1 R2 C1 C2 Vi Vo t $\Leftrightarrow$ \\
$\mathtt{\ }$\hspace{2.25cm}  diff\_eq\_n\_order 1 (outlst\_pid\_contr R1 C2) Vo t = \\
$\mathtt{\ }$\hspace{2.25cm}  diff\_eq\_n\_order 2 (inlst\_pid\_contr R1 R2 C1 C2) Vi t }}}
\end{definition}
\end{flushleft}

We verified the behavioural specification based on the implementation of the PID controller as the following theorem:

	\begin{flushleft}
		\begin{theorem}
			\label{THM:behav_imp_implem}
	{\footnotesize
				\textup{\texttt{$\vdash$ $\forall$ R1 R2 C1 C2 Vi Va Vb Vo t. \&0 < R1 $\wedge$ \&0 < R2 $\wedge$   \\
						$\mathtt{\ }$\hspace{0.0cm} \&0 < C1 $\wedge$ \&0 < C2 $\wedge$ ($\forall$t. differentiable\_higher\_derivative Vi Vo Vb t) $\wedge$  \\
						$\mathtt{\ }$\hspace{0.4cm} pid\_controller\_implem Vi Vo Va Vb C1 C2 R1 R2  \\
			$\mathtt{\ }$\hspace{0.5cm} $\Rightarrow$ (\&0 < drop t $\Rightarrow$ pid\_controller\_behav\_spec R1 R2 C1 C2 Vi Vo t)
			}}}
		\end{theorem}
	\end{flushleft}

\noindent The first four assumptions model the design requirement for the underlying system. The next assumption provides the differentiability of the higher-order derivatives of $\texttt{Vi}$, $\texttt{Vo}$ and $\texttt{Vb}$ up to the order 1, 2 and 2, respectively. The last assumption presents the implementation for the PID controller. Finally, the conclusion presents its behavioral specification. We also develop a simplification tactic $\texttt{DIFF\_SIMP\_TAC}$, which simplifies the behavioural specifications of the PID controller as well as the other controllers and compensators~\cite{adnan2016facstp_wp}.

Next, we verified the transfer function of the PID controller as follows:

	\begin{flushleft}
		\begin{theorem}
			\label{THM:implem_imp_tf_model_pid}
		{\footnotesize
				\textup{\texttt{$\vdash$ $\forall$ R1 R2 C1 C2 Vi Vo s t. \&0 < R1 $\wedge$ \&0 < R2 $\wedge$ \&0 < C1 $\wedge$ \\
						$\mathtt{\ }$\hspace{0.0cm} $\sim$(laplace\_transform Vi s = Cx (\&0)) $\wedge$ $\sim$(Cx R1 $\ast$ Cx C2 $\ast$ s = Cx (\&0)) $\wedge$ \\
						$\mathtt{\ }$\hspace{0.0cm} \&0 < C2 $\wedge$ ($\forall$t. differentiable\_higher\_derivative Vi Vo t) $\wedge$  \\
                        $\mathtt{\ }$\hspace{0.0cm} laplace\_exists\_higher\_deriv Vi Vo s $\wedge$ zero\_initial\_conditions Vi Vo  $\wedge$ \\
						$\mathtt{\ }$\hspace{0.0cm} ($\forall$t. pid\_controller\_behav\_spec R1 R2 C1 C2 Vi Vo t)   \\
						$\mathtt{\ }$\hspace{1.0cm} $\Rightarrow$ $\mathtt{\dfrac{laplace\_transform\ Vo\ s}{laplace\_transform\ Vi\ s}}$ =       $\mathtt{\dfrac{\texttt{--}\big(Cx (R1 \ast C1 \ast R2 \ast C2) \ast s\ pow\ 2 + \left(Cx (R2 \ast C2) + Cx (C1 \ast R1)\right) \ast s + Cx (\&1)\big)}{Cx (R1 \ast C2) \ast s}}$
			}}}
		\end{theorem}
	\end{flushleft}

\noindent The first six assumptions present the design requirements for the underlying system. The next two assumptions provide the differentiability and the Laplace existence condition for the higher-order derivatives of $\texttt{Vi}$ and $\texttt{Vo}$ up to the order $2$ and $1$, respectively. The next assumption models the \textit{zero initial conditions} for the voltage functions $\texttt{Vi}$ and $\texttt{Vo}$. The last assumption presents the behavioural specification of the PID controller.
Finally, the conclusion of Theorem~\ref{THM:implem_imp_tf_model_pid} presents its required transfer function.
By judicious selection of the configuration of passive components, we obtain various controllers, such as P, I, D, PI, PD and perform the above-mentioned analysis for all of them.

Compensators are widely used in control systems, to improve their frequency response, steady-state error and the stability and hence, act as a fundamental block of a control system. Like controllers, the compensators are also modeled using their active realizations. A compensator uses the same analog circuit, which is used for the controllers, presented in Fig.~\ref{FIG:generic_active_realization}, by making an appropriate choice of the passive components $C_A$ and $C_B$, as shown in Fig.~\ref{FIG:config_compensator}. It acts as a lag-compensator under the condition $ R_2C_2 > R_1C_1$, whereas for the case of $ R_1C_1 > R_2C_2$, it acts as a lead-compensator.
The configurations of the passive components for the controllers and compensators, and their formalization is presented in~\cite{adnan2016facstp_wp}.

Compensators are also modeled using their passive realizations based on an electrical circuit, which comprises of two components, i.e., $C_A$ and $C_B$, which are shown as rectangular boxes in
Fig.~\ref{FIG:generic_passive_realization}.
The boxes $C_A$ and $C_B$ contain different configurations of the passive components, i.e., resistors and capacitors. By making an appropriate choice of these passive components, we obtain various compensators, such as lag, lead and lag-lead~\cite{nise2007control}.
The configuration of the lag-lead compensator is shown in Fig.~\ref{FIG:config_lead_lag_compensator}. Moreover, the configurations of the passive components for the compensators and their formalization in HOL Light is presented in~\cite{adnan2016facstp_wp}.

The formalization of this section took around 300 lines of HOL Light code and around 14 man-hours. This clearly illustrates the effectiveness of our foundational formalization, presented in the previous section.

\section{Unmanned Free-Swimming Submersible Vehicle} \label{SEC:unmanned_vehicle}

Unmanned Free-Swimming Submersible (UFSS) vehicles are a kind of autonomous underwater vehicles (AUVs) that are used to perform different tasks and operations in the submerged areas of the water. These vehicles have their own power and control systems, which are autonomously operated and controlled by the onboard computer system without any involvement of human assistance as it is difficult for humans to work in an underwater environment.
UFSS vehicles are used in many safety-critical domains to perform different tasks, such as underwater navigation and object detection~\cite{kondo2004navigation}, performing deep sea rescue and salvage operations~\cite{wernli2000low}, searching for sea mines~\cite{willcox2001bluefin} and securing sea harbour~\cite{willcox2001bluefin}.
Due to their wider usage in the above-mentioned safety-critical applications, an accurate analysis of their control system is of utmost importance.

\begin{figure}[H]
\centering
\scalebox{0.23}
{\includegraphics[trim={4 0.3cm 4 0.3cm},clip]{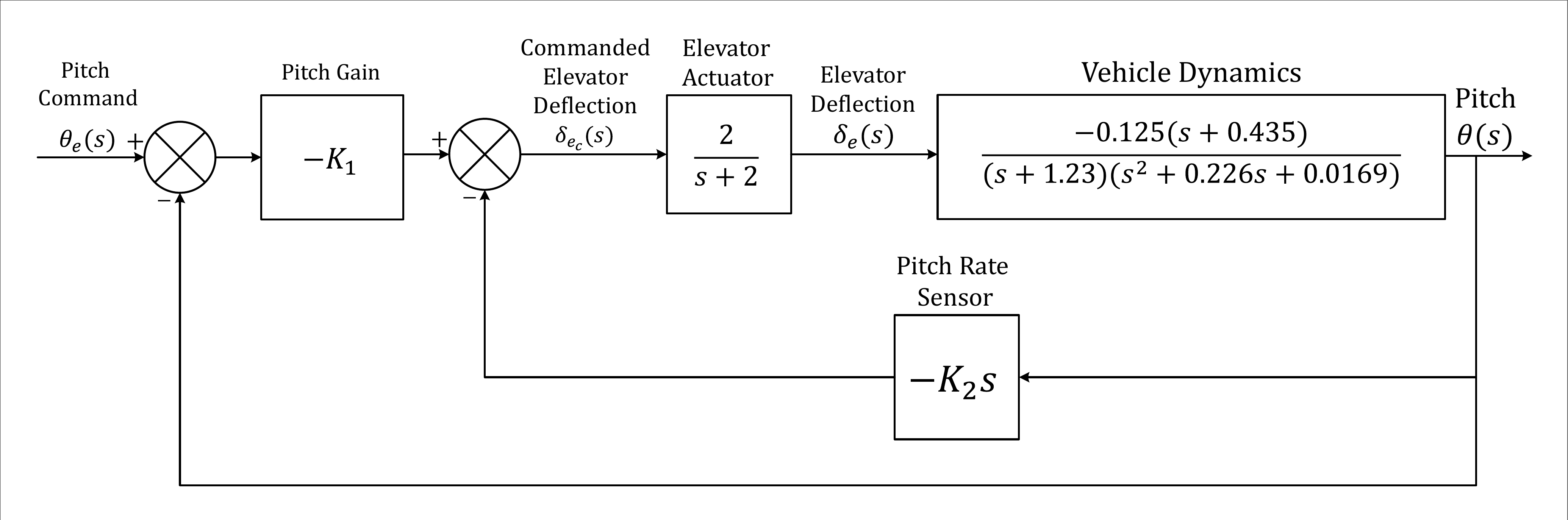}}
\caption{Pitch Control Model for Unmanned Free-swimming Submersible Vehicle}
\label{FIG:unmanned_vehicle}
\end{figure}

We present a formal analysis of the pitch control system of a UFSS vehicle. The pitch control system is responsible for the uninterrupted operation and functionality of the UFSS vehicle by manipulating different parameters, such as, elevator surface, pitch angle~\cite{nise2007control}.
Fig.~\ref{FIG:unmanned_vehicle} depicts its block diagram.

\vspace{-0.5cm}

\vspace{-0.5cm}

The dynamics of the UFSS vehicle are represented by its corresponding differential equation, which presents the relationship between the pitch command angle $\theta_e (t)$ and the pitch angle $\theta(t)$, and is given as follows:

\small

\begin{equation}\label{EQ:diff_eq_unmanned_vehicle}
\begin{split}
\dfrac{d^4\theta}{{dt}^4} + 3.456\dfrac{d^3\theta}{{dt}^3} + (3.207 + 0.25 K_2)\dfrac{d^2\theta}{{dt}^2} + (0.616 + 0.1088 K_2 + 0.25 K_1)\dfrac{d\theta}{dt} + \\
(0.1088 K_1 + 0.0416) = 0.25 K_1\dfrac{d\theta_e}{dt} + 0.1088 K_1
\end{split}
\end{equation}

\normalsize

We formalize the above differential equation as follows~\cite{rashid2016formalization}:


\begin{flushleft}
\begin{definition}
\label{DEF:diff_eq_unmanned_vehicle}
{\footnotesize
\textup{\texttt{\hspace{-0.1cm}$\vdash$ \footnotesize{$\forall$ K1. \hspace{-0.3cm} inlst\_ufsv K1 \hspace{-0.2cm} =  \hspace{-0.2cm} $\mathtt{\left[Cx \left(\#0.1088\right) \ast Cx\ K1; Cx \left(\#0.25\right)\ast Cx\ K1\right]}$}
}}} \\
{\footnotesize
\textup{\texttt{$\vdash$ \footnotesize{ $\forall$ K1 K2. outlst\_ufsv K1 K2 = $\mathtt{\big[Cx \left(\#0.1088\right) \ast Cx\ K1 + Cx \left(\#0.0416\right); Cx \left(\#0.25\right) \ast Cx\ K1\ + Cx \left(\#0.1088\right) \ast Cx\ K2}$} \\
$\mathtt{\ }$\hspace{-0.3cm}  $\mathtt{+ \ Cx \left(\#0.6106\right); Cx \left(\#0.25\right) \ast Cx\ K2 + Cx\left(\#3.207\right); Cx\left(\#3.456\right); Cx\ (\&1) \big]}$
}}} \\
{\footnotesize
\textup{\texttt{\footnotesize{$\vdash$ diff\_eq\_ufsv inlst\_ufsv outlst\_ufsv theta thetae K1 K2 $\Leftrightarrow$ \\
$\mathtt{\ }$\hspace{2.25cm} ($\forall$t. diff\_eq\_n\_order 4 (outlst\_ufsv K1 K2) theta t = \\
$\mathtt{\ }$\hspace{3.10cm} diff\_eq\_n\_order 1 (inlst\_ufsv K1) thetae t) }}}}
\end{definition}
\end{flushleft}

\noindent where $\texttt{thetae}$ and $\texttt{theta}$ represent the input and the output of the pitch control system and $\texttt{K1}$ and $\texttt{K2}$ are the pitch gain and pitch rate sensor gain, respectively. The symbol $\mathtt{\#}$ is used to represent a decimal number of data type $\mathds{R}$ in HOL Light and is same as symbol $\mathtt{\&}$ for the integer literal of data type $\mathds{R}$.

The transfer function of the pitch control of the UFSS vehicle is as follows:

\small

\begin{equation}\label{EQ:transfer_fun_unmanned_vehicle}
\begin{split}
\dfrac{\theta (s)}{\theta_e(s)} = \dfrac{0.25K_1 s + 0.1088 K_1}{\splitdfrac{s^4 + 3.456 s^3 + (3.207 + 0.25K_2)s^2 + (0.6106 + 0.1088K_2 +}{ 0.25 K_1)s + (0.1088 K_1 + 0.0416)}}
\end{split}
\end{equation}

\normalsize

We verified the above transfer function as the following HOL Light theorem:

\begin{flushleft}
\begin{theorem}
\label{THM:transfer_fun_unmanned_vehicle}
{\footnotesize
\textup{\texttt{$\vdash$ $\forall$ thetae theta s K1 K2.   \\
	$\mathtt{\ }$\hspace{-0.1cm} ($\forall$t. differentiable\_higher\_deriv theta thetae t) $\wedge$   \\
	$\mathtt{\ }$\hspace{-0.1cm} laplace\_exists\_of\_higher\_deriv theta thetae s $\wedge$   \\
	$\mathtt{\ }$\hspace{-0.1cm} zero\_init\_conditions theta thetae $\wedge$  \\
	$\mathtt{\ }$\hspace{-0.1cm} diff\_eq\_ufsv inlst\_ufsv outlst\_ufsv theta thetae K1 K2 $\wedge$  \\
	$\mathtt{\ }$\hspace{-0.1cm} non\_zero\_denominator\_condition theta s \\
 $\mathtt{\ }$\hspace{1.0cm} \vspace{-0.2cm} \\
	$\mathtt{\ }$\hspace{1.0cm} $\Rightarrow$ $\mathtt{\dfrac{laplace\_transform\ theta\ s}{laplace\_transform\ thetae\ s}}$ = \\
 $\mathtt{\ }$\hspace{1.0cm} \vspace{0.2cm} \\
	$\mathtt{\ }$\hspace{-0.3cm} \footnotesize{$\mathbf{\mathtt{\dfrac{\left(Cx \left(\#0.25\right) \ast Cx\ K1\right) \ast s + Cx \left(\#0.1088\right)  \ast Cx\ K1}{\splitdfrac{s\ pow\ 4 + Cx \left(\#3.456\right) \ast s\ pow\ 3 + \Big(Cx \left(\#0.25\right) \ast Cx\ K2 + Cx \left(\#3.207\right)\Big) }{\splitdfrac{\ast\ s\ pow\ 2\ + \Big(Cx \left(\#0.25\right) \ast Cx\ K1 + Cx \left(\#0.1088\right) \ast Cx\ K2 + Cx \left(\#0.6106\right)\Big) } {\ast s\ +  Cx \left(\#0.1088\right) \ast Cx\ K1 + Cx \left(\#0.0416\right) }   }}}}$}
}}}
\end{theorem}
\end{flushleft}

The first two assumptions present the differentiability and the Laplace existence condition of the higher-order derivatives of $\texttt{thetae}$ and $\texttt{theta}$ up to order $1$ and $4$, respectively. The next assumption provides the \textit{zero initial conditions} for $\texttt{thetae}$ and $\texttt{theta}$. The next assumption  presents the differential equation specification for the pitch control system of UFSS vehicle. The final assumption models the non-negativity of the denominator of the transfer function presented in the conclusion of the above theorem. We also verified the open loop transfer function $\mathtt{\theta(s)/\delta_e(s)}$, frequency response (open and closed loop) and gain margin, for the UFSS vehicle and the details can be found in~\cite{adnan2016facstp_wp}.

The distinguishing feature of Theorem~\ref{THM:transfer_fun_unmanned_vehicle} and the other properties, compared to traditional analysis methods is their generic nature, i.e., all of the variables and functions are universally quantified and can thus be specialized in order to obtain the results for some given values. Moreover, all of the required assumptions are guaranteed to be explicitly mentioned along with the theorems due to the inherent soundness of the theorem proving approach. The high expressiveness of the higher-order logic enables us to model the differential equation and the corresponding transfer function in their true continuous form, whereas, in model checking they are mostly discretized and modeled using a state-transition system, which compromises the accuracy of the analysis.

To facilitate control engineers in using our formalization, we developed an automatic tactic $\texttt{\footnotesize{TRANSFER\_FUN\_TAC}}$, which automatically verifies the transfer function of the systems up to $20^{th}$-order. This tactic was successfully used for the automatic verification of the transfer functions of the controllers, compensators and the pitch control system of the UFSS vehicle. This automatic verification tactic only requires the differential equation and the transfer function of the underlying system and automatically verifies the transfer function. Thus, the formal analysis of the UFSS vehicle took only 25 lines of code and about half an hour, thanks to our automatic tactic and the foundational formalization of Section~\ref{SEC:Formalization_of_Laplace}.

\vspace{-0.3cm}

\section{Conclusion}\label{SEC:Conclusion}

\vspace{-0.3cm}

This paper presented a higher-order-logic theorem proving based approach for the formal analysis of the dynamical aspects of linear control systems using theorem proving. The main idea behind the proposed framework is to use a formalization of Laplace transform theory in higher-order logic to formally analyze the dynamic aspects of linear control systems.
For this purpose, we develop a new formalization of Laplace transform theory, which includes its formal definition and verification of its properties, such as linearity, frequency shifting, differentiation and integration in time domain, time shifting, time scaling, cosine and sine-based modulation and the Laplace transform of an $n$-order differential equation, which are used for the verification of the transfer function of a generic $n$-order linear control system.
Moreover, the paper also presents the formal verification of some widely used linear control system characteristics, such as frequency response, phase margin and the gain margin, using the verified transfer function, which can be used for the stability analysis of a linear control system. We also formalize the active realization of various controllers, such as PID, PD, PI, P, I, D, and various compensators, such as lag and lead. Finally, we formalize the passive realization of the various compensators, such as lag, lead and lag-lead and verified the corresponding behavioral (differential equation) and the transfer function specifications.
To facilitate the usage of these formalizations in analyzing real-world linear control systems, we developed some simplification and automatic verification tactics, in particular the tactic $\texttt{\small{TRANSFER\_FUN\_TAC}}$, which automatically verifies the transfer function of any real-world linear control system based on its differential equation.
These foundations can be used to analyze a wide range of linear control systems and for illustration purposes, the paper presents the formal analysis of an unmanned free-swimming submersible vehicle.

In future, we plan to link the proposed formalization with Simulink so that the users can provide the system model as a block diagram. This diagram can be used to extract the corresponding transfer function~\cite{babuska1999matlab}, which can in turn be formally verified, almost automatically, to be equivalent to the corresponding block diagram based on the reported formalization and reasoning support.

\vspace{-0.3cm}

\section*{Acknowledgements}

\vspace{-0.3cm}

This work was supported by the National Research Program for Universities grant (number 1543) of Higher Education Commission (HEC), Pakistan.

\bibliographystyle{splncs03}
\bibliography{bibliotex}

\end{document}